%% file: main.tex
\documentclass[%
 reprint,
 amsmath,amssymb,
 aps,
]{revtex4-2}

\usepackage{tikz}
\usepackage{pgfplots}
\usepackage{graphicx}
\usepackage[outline]{contour}
\usepackage{xcolor}
\usepackage{dcolumn}
\usepackage{bm}

\begin{document}

\preprint{APS/123-QED}

\title{Manipulating quasi-bound states in a photonic crystal with \\ periodic impurities to store quantum information}

\author{Benjamin Rempfer}
 \email{brempfer99@gmail.com}
 \affiliation{Mathematical Sciences Department, Butler University}
 \altaffiliation{Current Affiliation: MIT Lincoln Laboratory}
\author{Gonzalo Ordonez}%
 \email{gordonez@butler.edu}
 \affiliation{Department of Physics and Astronomy, Butler University}%

\begin{abstract}
    We analytically model a one-dimensional lattice with periodic impurities representing a photonic crystal from first principles. We then investigate bound states in the continuum by computing the transmission and reflection coefficients. It turns out that when there are more impurities in our designed system then there exists a wider range of wavenumbers where particles become essentially trapped. A perturbative-based explanation is shown to verify this phenomenon quantitatively. Due to this window of wavenumbers quantum information could be encoded in our system by constructing differently shaped wave packets that are bound by the tuning of parameters in our system.
\end{abstract}
\maketitle

\section{\label{sec:level1}Introduction and Motivation}

The phenomenon of electron trapping in semiconductor quantum wires has been extensively studied in the past several decades in no small part due to its application to quantum computation and quantum information storage \cite{tanaka, perez}. However, we assert that it is possible to achieve comparable results in regards to information storage with wave packets of photons propagating through a photonic crystal instead of an electron in a semiconductor or similar quantum dot techniques \cite{2014}. This article lays the ground work for future investigations on the potential storage capacity of photonic systems in quantum devices. With technology in its infancy due to various unsolved experimental and engineering problems such as decoherence and noise there is much room for research into possible quantum information storage schemes \cite{storage}.

Both theoretical and experimental research has shown it possible to effectively `slow' light within a photonic crystal arrangement by utilizing electromagnetically induced transparency and other optical techniques \cite{goldzak, yanik, hau}. In other words, the group velocity of a wave packet of light is slowed to speeds on the order of $10^{-7}$ that of $c$ in a vacuum. Still, there is not much work done to explain the trapping of photons in terms of fundamental quantum mechanics. Thus, we will attempt to work from first principles, while building off of recent work concerning bound states in the continuum \cite{tanaka}. In this work, it was shown that a particle with a specific wavelength could be trapped in a bound-state in continuum,  between two impurities attached to a one-dimensional lattice. The first problem that arises is how to design a system that affects more than one wave number of light. To encode information one would need a spectrum of wavenumbers.

How can we guarantee that more than a single photon is bound in our system? Before delving into the answer to this question it is necessary to discuss how one can model a photonic crystal quantum mechanically. The process is akin to coupling together lattice sites that correspond to separate quantum states \cite{feynman}. For simplicity we will consider a one-dimensional lattice devoid of interactions with its environment. In practice, these states can take several shapes such as parallel wave guides or various crystalline atomic structures \cite{chien, silicon, intro}. The appropriate parameters such as the bond energy between states and the energy of each quantum state ensure each excitation of our model is accurate. 

\begin{figure*}
\input{fig.tex}
\caption{\label{fig:wide}Diagram of our $N$--impurity system with all relevant nomenclature included.}
\end{figure*}
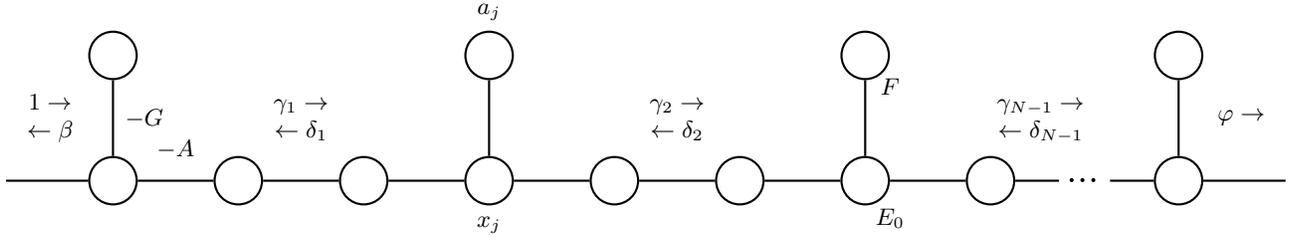

An arbitrary number $N$ of impurities attached to this one-dimensional array of quantum sites allows the effective trapping of more than one wave number, in fact a range of wave numbers. The entirety of this paper is dedicated to rigorously demonstrating this fact. In order to analyze this model quantitatively we must begin with the famous Schr\"{o}dinger equation. But, due to the magnitude of our system we end up with a $4N$ -- dimensional system of partial differential equations. 

Let us assume that each binding between quantum sites allows for a certain complex amplitude transfer in either direction. With $E_0$ given as the energy of the ``pure'' states, $-A$ as the bond energy between them, and $x_m(t)$ as the wave amplitude value at some pure index $m$, the amplitudes at pure sites not connected to impurities obey the differential equation
\begin{eqnarray*}
i\hbar \frac{dx_m(t)}{dt}=E_0x_m(t)-Ax_{m-1}(t)-Ax_{m+1}(t),
\end{eqnarray*}
whereas amplitudes at sites connected to impurities obey the equation
\begin{eqnarray*}
i\hbar \frac{dx_m(t)}{dt}=E_0x_m(t)-Ax_{m-1}(t)-Ax_{m+1}(t)-G a_m(t),
\end{eqnarray*}
with $a_m(t)$ denoting the amplitude at the coupled impurity and $-G$ denoting the bond energy between \textit{pure} and \textit{impure} states of equal index. Finally, impurity amplitudes with onsite energy $F$ obey the equation
\begin{eqnarray*}
i\hbar \frac{da_m(t)}{dt}=Fa_m(t)-Gx_{m}(t).
\end{eqnarray*}
Hereafter we will consider regularly spaced impurities, attached at sites $m=nj$ for some fixed integer $j>1$ and $n\in\mathbb{Z}^+_{N}$.

Now, in order to solve this arbitrarily long list of coupled differential equations we must employ a test solution. This test solution is commonly utilized when dealing with amplitudes that change simultaneously with identical frequencies \cite{feynman}. This trial solution is $x_m(t)=x_me^{-iEt/\hbar}$ where $x_m$ is some constant complex amplitude and similarly $a_m(t)= a_m e^{-iEt/\hbar}$. We will, from now on, denote these amplitudes as $\beta$ for the wave reflected by the first $0\mbox{th}$ \textit{impurity} at site $0$, $\gamma_n$ for the wave transmitted from the $(n-1)\mbox{th}$ \textit{impurity}, $\delta_n$ for the wave reflected by the $n\mbox{th}$ impurity, and $\varphi$ for the wave transmitted from the final $(N-1)\mbox{th}$ \textit{impurity}. For sites not connected to impurities, we will assume $x_m=\gamma_n e^{ikbm} + \delta_n e^{-ikbm}$ (with a given wave number $k$ and coefficients $\gamma_n$, $\delta_n$ to be determined), which gives $E-E_0=-2A\cos{\left(kb\right)}$. Next, an amended subset of the $4N$  equations is given. Set $\Delta E=E-E_0$.
\begin{align*}
&\Delta E(\gamma_n e^{ikb(nj-1)}+\delta_n e^{-ikb(nj-1)}) \\ 
&\hspace{15pt}=-A(x_{nj}+\gamma_n e^{ikb(nj-2)}+\delta_n e^{-ikb(nj-2)}) \\
&\Delta Ex_{nj}=-A(\gamma_ne^{ikb(nj-1)}+\delta_ne^{-ikb(nj-1)} \\
&\hspace{15pt}+\gamma_{n+1}e^{ikb(nj+1)}+\delta_{n+1}e^{-ikb(nj+1)})-Ga_{nj} \\
&(E-F)a_{nj}=-Gx_{nj} \\
&\Delta E(\gamma_{n+1}e^{ikb(nj+1)}+\delta_{n+1}e^{-ikb(nj+1)}) \\
&\hspace{15pt}=-A(\gamma_{n+1}e^{ikb(nj+1)}+\delta_{n+1}e^{-ikb(nj+2)}+x_{nj})
\end{align*}
The four above equations deal with the Schr\"{o}dinger equations describing the wave amplitudes at the site connected to the $n\mbox{th}$ impurity and its neighbors. For the zeroth impurity we assume $\gamma_0=1$ and denote $\delta_0 = \beta$. For the $(N-1)\mbox{th}$ impurity we assume $\delta_{N}=0$ and denote $\gamma_{N}=\varphi$. Figure 1 illustrates this system.

\section{\label{sec:level2}Quasi-Bound States for an $N$ Impurity System}

Now that we have established the required Schr\"{o}dinger equations we must solve them for the final transmission amplitude $\varphi$ as a function of wave number $k$, the number of impurities $N$ and the reflection amplitude $\beta$. From the systems of $4$ equations above we eliminate $a_{nj}$ and $x_{nj}$ and solve for $(\gamma_{n+1}, \delta_{n+1})$ in terms of $(\gamma_{n}, \delta_{n})$. Iterating this procedure we finally obtain an equation relating $\varphi$ to  $\beta$ as follows.
\begin{equation}
\begin{pmatrix}
\varphi \\
0
\end{pmatrix} = U^{N-1}(TU^{-1})^{N}U \begin{pmatrix} 1 \\ \beta \end{pmatrix} 
\end{equation}
where
\begin{equation*}
U =\begin{pmatrix}
e^{-ikb} & 0 \\
0 & e^{ikb}
\end{pmatrix},
\end{equation*}
\begin{equation*}
T=
\begin{pmatrix}
1+X & X \\
-X & 1-X
\end{pmatrix},
\end{equation*}
and
\begin{equation*}
X = \frac{G^2}{A(E-F)2i\sin(kb)}.
\end{equation*}
Before proceeding we shall pause to note that our goal is to attain a quasi-bound state in the continuum in which photons are trapped in the section of the lattice where the impurities are attached. To achieve this,  we must create a quantum resonance by setting the energy of the impurities to a special value $F=E_0-2A\cos{(m\pi/j)}$ \cite{tanaka}. Similarly, the wavenumber of the exactly trapped light is then $k=\frac{m\pi}{bj}$ with $m\in\mathbb{Z}$. This bound state in continuum is characterized by the vanishing of the transmission probability $|\varphi|^2$, which implies a $100\%$ reflection probability.

One may be left with a natural set of curiosities relating to bound states in the continuum, quantum resonance, and trapping light. These are best encapsulated in the following question: why does a transmission probability of $0$ imply a bound state in the continuum? When a one-dimensional wave is incident on a discrete state that may resonate with the wave, the reflection probability  $R$ is approximately given by
 \begin{align*} 
 R \approx \frac{\Gamma^2}{(E_k-E_r)^2 + \Gamma^2}
 \end{align*}
 where $E_k$ is the energy of the wave, $E_r$ is the energy of the discrete state and $\Gamma$ is the width of the resonance. At the exact resonance point, $E_k=E_r$, the reflection probability is $1$ and thus the transmission probability is $0$. The bound-state in the continuum is a special type of resonant state with vanishing width, $\Gamma \to 0$ \cite{fan}. In this limit $R \to 0$ everywhere, except at the exact resonance point, where $R = 1$. The vanishing $\Gamma$ gives the resonance an infinite lifetime, which allows a photon to remain permanently trapped.
 
Another way to see this is that the bound state in continuum results from the interference of waves scattered off each impurity. In the regions away from the impurity, there is destructive interference, while in the impurity region there is  constructive interference. This results in a standing waves in the impurity region, where a single photon is effectively trapped.  

What makes this system interesting and novel is that the transmission probability can be made arbitrarily small for a window of wavenumbers centered around $k=\frac{m\pi}{bj}$, as seen in Fig. 2. We will discuss the width of this box dip in the next section.

\begin{figure}[b]
\includegraphics[width=0.85\linewidth]{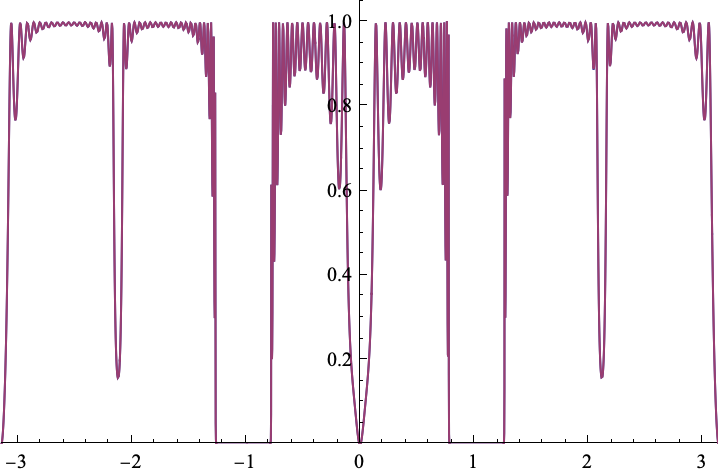}
\caption{\label{fig:epsart} The graph of transmission probability $|\varphi|^2$ versus the wavenumber $k$ with the wavenumber varied as $-\pi\leq k\leq\pi$. The transmission probability dips to zero around $k=m\pi/bj$, with $b=1$, $m=\pm 1$ and $j=3$. With these parameters the dips occur around $k=\pm\pi/3$. Other parameters are $N=22$, $E_0=0$, $A=1$, $G=0.5$, and $F=-2A \cos(m\pi/j)$. } 
\end{figure}

\section{\label{sec:leve3}Perturbative Approach}

It is clear graphically that there is a range of wavenumbers that correspond to ``quasi-bound" states (see Figures 2 and 3). If an experimentalist sends a wave packet through this system and concurrently tunes the parameters to achieve resonance then quantum information would be stored between the end impurities. 

\begin{figure}[b]
\includegraphics[width=0.9\linewidth]{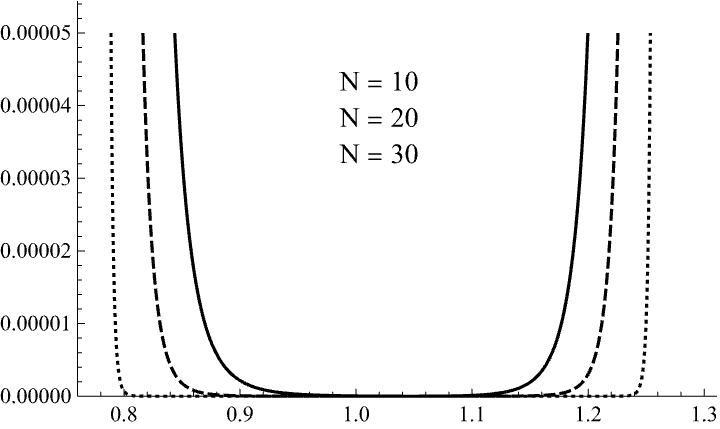}
\caption{\label{fig:epsart} The graph of transmission probability $|\varphi|^2$ versus the wavenumber $k$ with the wavenumber zoomed in around a neighborhood of $k=\frac{m\pi}{bj}=\pi/3$. The width of the box dip increases as the number of impurities $N$ increases.}
\end{figure}

Another (analytic) way to understand the behavior of each wavenumber and why there is a range that corresponds to `quasi-bound' states is through perturbation. This technique is common in quantum mechanics but slightly differs here due to our ability to find an exact solution as evidenced in the previous section. We will start with equation (1) and work towards a single equation involving $|\varphi|^2$, $N$, and $k$. 
We denote $H=TU^{-1}$ and the left eigenvectors of $H$ as $\left(\psi_{i_1}\,\,\psi_{i_2}\right)$, with eigenvalues $\lambda_i$. Multiplying by one of these eigenvectors we get
\begin{equation*}
\underset{}{\left(\psi_{i1}\,\,\psi_{i2}\right) U^{1-N}
\begin{pmatrix}
\varphi \\
0
\end{pmatrix}}
=
\lambda_i^{N}\left(\psi_{i1}\,\,\psi_{i2}\right) U
\begin{pmatrix}
1 \\
\beta
\end{pmatrix}
\end{equation*}
Since the determinant of $H$ is $1$, the product of its eigenvalues must also be $1$. Thus we write $\lambda_1=\lambda,\lambda_2=\lambda^{-1}$ and obtain the following system of equations:

\begin{align*}
e^{ikbj(N-1)} \psi_{11} \varphi &= \lambda^{N}\left(e^{-ikbj}\psi_{11}+ e^{ikbj}\beta\psi_{12}\right) \\
e^{ikbj(N-1)} \psi_{21} \varphi &= \lambda^{-N}\left(e^{-ikbj}\psi_{21}+ e^{ikbj}\beta\psi_{22}\right)
\end{align*}
which gives the transmission amplitude 
\begin{equation*}
\varphi=e^{-ikbjN}\frac{\psi_{12}\psi_{21}-\psi_{11}\psi_{22}}{\psi_{12}\psi_{21} \lambda^{N}-\psi_{11}\psi_{22}\lambda^{-N}}.
\end{equation*}
The eigenvalue equations yield
\begin{align*}
    \left(\psi_{11}\,\,\psi_{12}\right) H &= \lambda \left(\psi_{11}\,\,\psi_{12}\right) \\
    \left(\psi_{21}\,\,\psi_{22}\right) H &= \lambda^{-1} \left(\psi_{21}\,\,\psi_{22}\right)
\end{align*}
and setting $y=e^{-ikbj}$ we get
\begin{align*}
    \psi_{12} &= \left(\frac{1+X}{X}-\frac{y}{X} \lambda\right) \psi_{11}\\
    \psi_{21} &= \left(\frac{1-X}{-X}+\frac{1}{yX} \lambda^{-1}\right) \psi_{22}
\end{align*}

which allows the final result of 
\begin{align*}
\varphi=&e^{-ikbjN}\left(-2+y\lambda+(y\lambda)^{-1}-\left(y\lambda-(y\lambda)^{-1}\right)X\right) \nonumber\\&\hspace{0.5pt}\{\left[X^2-2+y\lambda+(y\lambda)^{-1}-\left(y\lambda-(y\lambda)^{-1}\right)X\right] \nonumber\\&\hspace{2cm}\lambda^{N}-\lambda^{-N}X^2\}^{-1}.
\end{align*}
The eigenvalue equations also give the eigenvalues $\lambda, \lambda^{-1}$ of $H$:
\begin{align*} 
 \lambda^{\pm 1} &= \cos(kbj) + iX\sin(kbj) \nonumber\\&\pm \sqrt{\left[\cos(kbj) + iX \sin(kbj)\right]^2-1}
 \end{align*}
A wide dip of the transmission probability $|\varphi|^2$ occurs for large $N$ around the point $k=\frac{m\pi}{bj}$ again with $m\in\mathbb{Z}$. We will  study the limit of $\varphi$ when $\epsilon$ tends to $0$ in $k = \frac{m\pi}{bj} +\epsilon$. In this limit
\[\lim_{\epsilon\to0}iX \sim \frac{\alpha}{\epsilon b j} = \infty, \mbox{\,\,\,where } \alpha= \frac{G^2 j}{4A^2\sin^2(m\pi/j)}\]
Furthermore, \[\lim_{\epsilon\to0} \lambda  = (-1)^m (1+\alpha) +  \sqrt{\left(1+\alpha\right)^2-1}
= (-1)^m e^{(-1)^m \gamma}\]
 where $\cosh(\gamma)= 1+ \alpha, \gamma>0$
and $\lim_{\epsilon\to0} y = (-1)^m$
 which give, for large $N$,
  \[\lim_{\epsilon\to0} \varphi \sim \epsilon \, e^{-N \gamma} \]
 
These limits explain the behavior of $\varphi$ as it is perturbed slightly off of its critical point precisely where the transmission probability is zero.  As the number $N$ of impurities increases, the transmission amplitude $\varphi$ decreases exponentially in the vicinity of the resonant wave number $k=\frac{m\pi}{bj}$, thereby increasing the width of the box dip. The behavior is consistent with that described in Section II.

\section{\label{sec:level4}Further Questions and Conclusion}

This paper presents a model of a one-dimensional lattice with periodic impurities that can, in principle, trap a range of wavenumbers. This spectrum is explored graphically as well as analytically. We explain how adding more impurities into our system yields a larger number of wavenumbers whose transmission probability asymptotically approaches $0$. Similarly, altering the impurity binding energy $G$ impacts the information storage capabilities of our system.
Our design and subsequent analysis of this one-dimensional system has lead to a series of natural questions. Answering these questions will extend the work of the authors and be of interest for many of the same reasons stated in the introduction. 
\begin{enumerate}
    \item Are certain implementations of photonic crystals more apt to trap light and/or store quantum information than others?
    \item Do other variations of our system widen the well of quasi-bound wavenumbers or approach zero asymptotically quicker than the way our system does as a function of the number of impurities $N$?
    \item Can we experimentally verify our results?
\end{enumerate}

\bibliography{main}
\end{document}

%% file: fig.tex
\begin{tikzpicture}

    \begin{scope}[every node/.style = {rectangle, thick, draw=none}]
    
        \node (N) at (-7.5,0) {\color{white}A};
        \node (O) at (10,0) {\color{white}A};
        \node (P) at (7.083,0) {\contour{black}{$\cdots$}};
    
        \node (Q) at (-6.667,0.833) {\begin{tabular}{c} 
                                     $1\rightarrow$ \\
                                     $\leftarrow\beta$
                                     \end{tabular}};
        \node (R) at (-3.333,0.833) {\begin{tabular}{c}
                                     $\gamma_1\rightarrow$ \\
                                     $\leftarrow\delta_1$
                                     \end{tabular}};
        \node (S) at (1.667,0.833) {\begin{tabular}{c}
                                    $\gamma_2\rightarrow$ \\
                                    $\leftarrow\delta_2$
                                    \end{tabular}};
        \node (T) at (6.500,0.833) {\begin{tabular}{c}
                                    $\gamma_{N-1}\rightarrow$ \\
                                    $\leftarrow\delta_{N-1}$
                                    \end{tabular}};
        \node (U) at (9.167,0.833) {$\varphi\rightarrow$};

        \node (V) at (-5.416,0.833) {$-G$};
        \node (W) at (-5,0.416) {$-A$};
        \node (X) at (4.500,1.250) {$F$};
        \node (Y) at (4.500,-0.500) {$E_0$};

        \node (Z) at (-0.833,-0.583) {$x_j$};
        \node (AA) at (-0.833,2.242) {$a_j$};
        
    \end{scope}

    \begin{scope}[every node/.style = {circle, thick, draw}]
    
        \node (A) at (-5.833,0) {\color{white}A};
        \node (B) at (-4.167,0) {\color{white}A};
        \node (C) at (-2.500,0) {\color{white}A};
        \node (D) at (-0.833,0) {\color{white}A};
        \node (E) at (0.833,0) {\color{white}A};
        \node (F) at (2.500,0) {\color{white}A};
        \node (G) at (4.167,0) {\color{white}A};
        \node (H) at (5.833,0) {\color{white}A};
        \node (I) at (8.333,0) {\color{white}A};

        \node (J) at (-5.833,1.667) {\color{white}A};
        \node (K) at (-0.833,1.667) {\color{white}A};
        \node (L) at (4.167,1.667) {\color{white}A};
        \node (M) at (8.333,1.667) {\color{white}A};
        
    \end{scope}

    \begin{scope}[every path/.style = {thick}]

        \path (N) edge (A);
        \path (O) edge (I);
    
        \path (A) edge (B);
        \path (B) edge (C);
        \path (C) edge (D);
        \path (D) edge (E);
        \path (E) edge (F);
        \path (F) edge (G);
        \path (G) edge (H);
        \path (H) edge (P);
        \path (P) edge (I);

        \path (A) edge (J);
        \path (D) edge (K);
        \path (G) edge (L);
        \path (I) edge (M);
    
    \end{scope}
    
\end{tikzpicture}